\pdfoutput=1
\documentclass[sigconf, natbib=false, screen, nonacm]{acmart}

\renewcommand\footnotetextcopyrightpermission[1]{} % removes footnote with conference information in first column

\copyrightyear{2022} 
\acmYear{2022} 
\setcopyright{acmcopyright}
\usepackage{amsthm}

\usepackage[T1]{fontenc}
\usepackage[utf8]{inputenc}

\usepackage{csquotes}
\usepackage{nicefrac}
\usepackage[textsize=tiny]{todonotes}
\usepackage{booktabs}
\usepackage[bookmarks=false]{hyperref}
\usepackage{tikz}
\usetikzlibrary{shapes.geometric, positioning, calc}
\usepackage{yquant}
\usepackage{physics}
\usepackage{csvsimple-l3}
\usepackage{pgfplotstable}

\usepackage{subcaption}
\usepackage[style=ieee, maxnames=6, minnames=1, date=year, doi=false, isbn=false, backend=biber, sortcites, dashed=false]{biblatex}
\usepackage[binary-units=true, detect-all=true]{siunitx}
\usepackage{etoolbox}
\robustify\bfseries

\usepackage{nicematrix}
\usepackage{balance}

\newtheorem{example}{Example}

\definecolor{RedEdge}{RGB}{191,40,40}
\definecolor{BlueEdge}{RGB}{40,191,191}
\tikzset{% 
	terminal/.style={draw,rectangle,inner sep=2pt,font=\footnotesize,very thick},
	zeronode/.style={fill, draw, circle, minimum width=2pt, inner sep=0pt,color=black},
	qubit/.style={draw,circle,inner sep=0pt,minimum width=0.35cm,minimum height=0.35cm,font=\footnotesize, thin},
	edgeOne/.style={color=RedEdge,ultra thick},
	edgeMOne/.style={color=BlueEdge,ultra thick},
	edgeSqrt/.style={color=RedEdge, thick},
	edgeMSqrt/.style={color=BlueEdge, thick},
}

\renewcommand{\baselinestretch}{0.98}

\begin{document}

\title[Handling Non-Unitaries in Quantum Circuit Equivalence Checking]{\huge Handling Non-Unitaries in Quantum Circuit Equivalence Checking}

\author[Lukas Burgholzer and Robert Wille]{\vspace*{-6mm}Lukas Burgholzer$^*$\hspace{3.0em}Robert Wille$^{*\dagger}$}
\affiliation{%
   \institution{$^*$Institute for Integrated Circuits, Johannes Kepler University Linz, Austria}
}
\affiliation{%
  \institution{$^\dagger$Software Competence Center Hagenberg GmbH (SCCH), Austria}
}
\email{{lukas.burgholzer, robert.wille}@jku.at}
\email{https://iic.jku.at/eda/research/quantum/}

\begin{abstract}
Quantum computers are reaching a level where interactions between classical and quantum computations can happen in real-time.
This marks the advent of a new, broader class of quantum circuits: \emph{dynamic quantum circuits}.
They offer a broader range of available computing primitives that lead to new challenges for design tasks such as simulation, compilation, and verification.
Due to the non-unitary nature of dynamic circuit primitives, most existing techniques and tools for these tasks are no longer applicable in an out-of-the-box fashion.
In this work, we discuss the resulting consequences for quantum circuit verification, specifically \emph{equivalence checking}, and propose two different schemes that eventually allow to treat the involved circuits as if they did not contain non-unitaries at all.
As a result, we demonstrate methodically, as well as, experimentally that existing techniques for verifying the equivalence of quantum circuits can be kept applicable for this broader class of circuits.
\end{abstract}

\maketitle

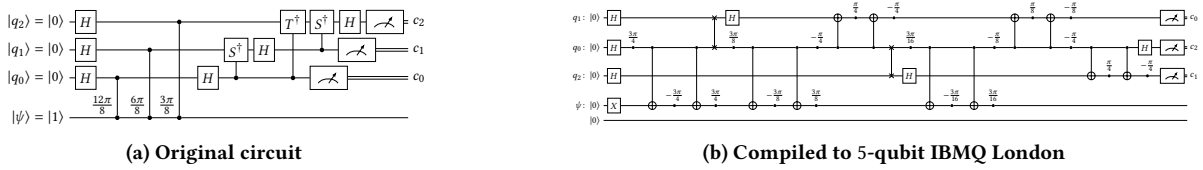
\begin{figure*}[t]
\centering	
\begin{subfigure}[b]{0.4\linewidth}
\centering
	\resizebox{0.8\linewidth}{!}{
\begin{tikzpicture}[label position=north west, every label/.style={inner sep=1pt}]
	\begin{yquant}
		qubit {$\ket{q_2} = \ket{0}$} q;
		qubit {$\ket{q_1} = \ket{0}$} q[+1];
		qubit {$\ket{q_0} = \ket{0}$} q[+1];
		qubit {$\ket{\psi} = \ket{1}$} q[+1];
		
		h q[0-2];
		[value=$\frac{12\pi}{8}$]
		phase q[3] | q[2];
		[value=$\frac{6\pi}{8}$]
		phase q[3] | q[1];
		[value=$\frac{3\pi}{8}$]
		phase q[3] | q[0];
		h q[2];
		box {$S^\dag$} q[1] | q[2];
		h q[1];
		box {$T^\dag$} q[0] | q[2];
		box {$S^\dag$} q[0] | q[1];
		h q[0];
%		align q[0-2];
		measure q[2];
		measure q[1];
		measure q[0];
		output {$c_0$} q[2];
		output {$c_1$} q[1];
		output {$c_2$} q[0];
	\end{yquant}
\end{tikzpicture}}
\caption{Original circuit}
\label{fig:qpe}
\end{subfigure}\hfill
\begin{subfigure}[b]{0.59\linewidth}
\centering
\resizebox{0.8\linewidth}{!}{
\begin{tikzpicture}
	\begin{yquant}
	qubit {$q_1\colon \ket{0}$} q;
	qubit {$q_0\colon \ket{0}$} q[+1];
	qubit {$q_2\colon \ket{0}$} q[+1];
	qubit {$\psi\colon \ket{0}$} q[+1];
	qubit {$\ket{0}$} q[+1];		
	h q[0];
	h q[1];
	h q[2];
	x q[3];
	phase {$\frac{3\pi}{4}$} q[1];
	cnot q[3] | q[1]; 
	phase {$-\frac{3\pi}{4}$} q[3]; 
	cnot q[3] | q[1]; 
	phase {$\frac{3\pi}{4}$} q[3];
	swap (q[0,1]); 
	h q[0]; 
	phase {$\frac{3\pi}{8}$} q[1]; 
	cnot q[3] | q[1]; 
	phase {$-\frac{3\pi}{8}$} q[3]; 
	cnot q[3] | q[1]; 
	phase {$\frac{3\pi}{8}$} q[3]; 
	phase {$-\frac{\pi}{4}$} q[1]; 
	cnot q[0] | q[1]; 
	phase {$\frac{\pi}{4}$} q[0]; 
	cnot q[0] | q[1]; 
	phase {$-\frac{\pi}{4}$} q[0];
	swap (q[1,2]); 
	h q[2];
	phase {$\frac{3\pi}{16}$} q[1]; 
	cnot q[3] | q[1]; 
	phase {$-\frac{3\pi}{16}$} q[3]; 
	cnot q[3] | q[1]; 
	phase {$\frac{3\pi}{16}$} q[3]; 
	phase {$-\frac{\pi}{8}$} q[1]; 
	cnot q[0] | q[1]; 
	phase {$\frac{\pi}{8}$} q[0]; 
	cnot q[0] | q[1]; 
	phase {$-\frac{\pi}{8}$} q[0];
	phase {$-\frac{\pi}{4}$} q[1]; 	
	cnot q[2] | q[1]; 
	phase {$\frac{\pi}{4}$} q[2]; 	
	cnot q[2] | q[1];
	phase {$-\frac{\pi}{4}$} q[2];
	h q[1]; 	
	align q[0-2];
	measure q[0];
	measure q[1];
	measure q[2];
	output {$c_0$} q[0];
	output {$c_1$} q[2];
	output {$c_2$} q[1];
	\end{yquant}
\end{tikzpicture}}
\caption{Compiled to $5$-qubit IBMQ London}
\label{fig:compiled_qpe}	
\end{subfigure}
\vspace*{-2mm}
\caption{$3$-bit precision QPE circuit for $U = \mathit{p}(\frac{3\pi}{8})$ and $\ket{\psi} = \ket{1}$, resulting in estimate $\tilde\theta = 0.c_2c_1c_0$}
\vspace*{-3mm}
\label{fig:qpecircs}
\end{figure*}

\section{Introduction}\label{sec:introduction}

Capabilities of quantum computers built today are steadily growing.
New devices do not only feature more and more qubits which are less prone to errors, but also allow for a much tighter classical control loop.
This is witnessed by the OpenQASM~$3.0$ specification recently published by IBM~\cite{crossOpenQASMBroaderDeeper2021} and the ability to perform conditional resets on IBM's quantum computers~\cite{ibmquantumQuantumCircuitsGet2021}.
Through the interaction of classical computation with the gates and measurements of a quantum circuit, new computing primitives such as \mbox{mid-circuit} measurements and resets as well as \mbox{classically-controlled} operations become possible within the coherence time for a single circuit execution.
We adopt the naming established by IBM and call this new, broader class of circuits \emph{dynamic quantum circuits}.

With these rapid advances in physical realizations comes the need for quantum software that aids developers and users to keep up with this pace.
Otherwise, we might end up in a situation where we have powerful quantum computers available, but no efficient means to use them.
Besides challenges, e.g., for classical/quantum design and compilation in general, this also poses new challenges for quantum circuit verification.

Verification of quantum circuits (more specifically, \emph{equivalence checking}) is an essential part in the modern quantum design flow.
To this end, the goal is to check whether two supposedly equivalent quantum circuits $G$ and $G'$ indeed realize the same functionality.
Important use cases include $(1)$ ensuring that the originally intended functionality of a quantum algorithm is preserved throughout the whole compilation process that the algorithm's circuit representation undergoes in order to be executable on an actual device, or $(2)$ ensuring that alternative (e.g., optimized) realizations of certain building blocks in quantum circuits are functionally equivalent to their original implementation.

In the past, several complementary approaches have been proposed for tackling this problem~\cite{yamashitaFastEquivalencecheckingQuantum2010, burgholzerAdvancedEquivalenceChecking2021,viamontesCheckingEquivalenceQuantum2007,niemannEquivalenceCheckingMultilevel2014,wangXQDDbasedVerificationMethod2008,smithQuantumComputationalCompiler2019, amyLargescaleFunctionalVerification2019, hongTensorNetworkBased2020}.
However, practically all of these approaches expect the underlying functionality to be unitary---which circuits containing dynamic circuit primitives no longer are.
As such, existing techniques for verifying conventional quantum circuits are not directly applicable in an out-of-the-box fashion.
In this work, we discuss the resulting consequences for quantum circuit equivalence checking and show that reinventing the wheel is not necessary in order to use existing tools for verifying this broader class of circuits.
To this end, we propose two different schemes targeted at two slightly different verification scenarios.

First, we consider the question whether two circuits $G$ and $G'$ which might contain dynamic circuit primitives are functionally equivalent as a whole. 
We show that, by combining well known results from quantum information, any such circuit can be \emph{transformed} to a circuit only containing unitary operations.
By transforming the dynamic circuit primitives in this fashion, existing techniques for checking the equivalence of quantum circuits can be employed for the broader class of dynamic circuits.

Second, we consider the question whether two circuits $G$ and $G'$ produce the same distribution of measurement outcomes given a fixed input state, i.e., whether they behave the same when executed on a quantum computer.
We show how to extract the complete measurement probabilities of a dynamic circuit, as if it did not contain non-unitaries, by cleverly applying classical quantum circuit simulation.

Experimental evaluations confirm that the proposed schemes indeed allow to handle the non-unitaries introduced by dynamic circuit primitives in an efficient fashion.
Overall, these schemes form a generic solution for handling non-unitaries in verifying the equivalence of quantum circuits that is applicable to any existing verification framework.

The rest of this work is structured as follows. 
\autoref{sec:background} provides the necessary background and motivation. 
Then, \autoref{sec:dyncirc} introduces dynamic circuits and explains the resulting problem for equivalence checking in detail, along with the general idea for solving this problem.
\autoref{sec:circuit_transform} and \autoref{sec:density_sim} elaborate on the proposed schemes and provide some discussion, while \autoref{sec:results} summarizes our experimental evaluations.
Finally, we conclude in \autoref{sec:conclusions}.

\section{Background and Motivation}\label{sec:background}
This section establishes the notation used in the remainder of this work and 
provides the necessary background information on quantum circuits.
We also review the \emph{Quantum Phase Estimation} algorithm (which is used as a running example) and motivate the importance of verifying quantum circuits. 
While the descriptions are kept brief, we refer the unacquainted reader to the provided references for further details.

\subsection{Quantum Circuits}\label{sec:qc}

In the traditional quantum circuit model~\cite{nielsenQuantumComputationQuantum2010,barencoElementaryGatesQuantum1995}, a quantum circuit $G$, acting on $n$ qubits, is specified by a sequence of $|G|$ quantum gates $g_0,\dots,g_{|G|-1}$.
Each quantum gate $g_i$, acting on $k\leq n$ qubits (most frequently $k=1$ or $k=2$), can be described by a $2^k\times 2^k$-dimensional unitary matrix $U_i$.

Given an initial state $\ket{\varphi}$ (represented as a $2^n$-dimensional state vector), the evolution of this initial state under the quantum circuit can be described by successively multiplying the individual gate matrices with the current state vector.
%\footnote{Before being applied to the state vector, the individual gate matrices have to be extended to the full system size by forming tensor products with identity matrices. We slightly abuse the notation and reuse the symbols $U_i$ also for the extended matrices whenever the meaning is apparent from the context.}.
Eventually, performing all multiplications results in a final state vector that encodes the probabilities of measuring the individual computational basis states.
When conducted on a classical computer, this is typically called \emph{(classical) quantum circuit simulation}.

\subsection{Quantum Phase Estimation}\label{sec:qpe}

The key ideas of this work will be illustrated by means of a particular quantum algorithm, namely \emph{Quantum Phase Estimation} (QPE,~\cite{nielsenQuantumComputationQuantum2010}),
which represents one of the key subroutines in important quantum algorithms such as Shor's algorithm~\cite{shorPolynomialtimeAlgorithmsPrime1997} for factoring numbers, the HHL algorithm~\cite{harrowQuantumAlgorithmLinear2009} for solving linear systems, or quantum principal component analysis~\cite{lloydQuantumPrincipalComponent2014} for machine learning.
It solves the problem of determining the phase of a unitary operator $U$ given an eigenstate $\ket{\psi}$, i.e., determining $\theta \in [0,1)$ such that $U \ket{\psi} = e^{2\pi i \theta} \ket{\psi}$.

To this end, the QPE algorithm determines an \mbox{$m$-bit} estimate \mbox{$\tilde\theta = 0.c_{m-1}\dots c_0$} of $\theta$.
First, controlled-$U^{2^k}$ operations ($0 \leq k < m$) are used to write the $m$-bit Fourier basis representation of $U$'s phase to an \mbox{$m$-qubit} register.
Afterwards, the inverse \emph{Quantum Fourier Transform} (QFT$^{\dag}$,~\cite{nielsenQuantumComputationQuantum2010}) is applied to transform the result to the computational basis.
Whenever~$\theta$ is representable using $m$ fractional bits, the algorithm succeeds with certainty, while otherwise, it yields a suitably high chance for success (with a probability larger than $\frac{4}{\pi^2}\approx 0.405$).

\begin{example}\label{ex:qpe}
Assume $U$ is given by $p(\frac{3\pi}{8})=\mathit{diag}(1, e^{2\pi i\frac{3}{16}})$ and $\ket{\psi} = \ket{1}$.
Then, \autoref{fig:qpe} shows the quantum circuit realizing the $3$-bit precision QPE algorithm.
It applies three rounds of controlled-phase rotations and then uses the three-qubit inverse Fourier transform to obtain the desired estimate $\tilde\theta = 0.c_2c_1c_0$ from the measurement results. Since $\theta = \frac{3}{16} = 0.0011_2$ cannot be exactly represented using three fractional bits, running the algorithm yields $\ket{001}$ and $\ket{010}$ as the most probable output states.
\end{example}

\subsection{Verification of Compilation Results}\label{sec:compflowver}

Executing a quantum algorithm on an actual quantum computer requires \emph{compiling} the algorithm's description $G$ to a representation $G'$ that adheres to all constraints imposed by the targeted device. This typically involves several steps such as synthesis~\cite{maslovAdvantagesUsingRelative2016,willeImprovingMappingReversible2013,degriendArchitectureawareSynthesisPhase2020}, mapping~\cite{siraichiQubitAllocation2018,zulehnerEfficientMethodologyMapping2019,willeMappingQuantumCircuits2019,liTacklingQubitMapping2019,sivarajahKetRetargetableCompiler2020}, and optimizations~\cite{itokoOptimizationQuantumCircuit2020,vidalUniversalQuantumCircuit2004,hietalaVerifiedOptimizerQuantum2019}.  

\begin{example}\label{ex:compiled}
Quantum computers manufactured by IBM natively support arbitrary single-qubit operations and the \mbox{two-qubit} controlled-NOT (or CNOT) operation.
A possible realization of the QPE circuit from \autoref{fig:qpe} on the \mbox{five-qubit}, \mbox{$T$-shaped} IBMQ London architecture is shown in \autoref{fig:compiled_qpe}. 
\end{example}

Verifying that the original circuit's functionality is preserved throughout the individual stages of the compilation process is a vital task in the quantum computing design flow.
In general, the functionality of a quantum circuit $G = g_0,\dots,g_{|G|-1}$ is represented by the $2^n\times 2^n$ system matrix $U = U_{|G|-1}\cdots U_0$.
Thus, comparing the functionality of two quantum circuits $G$ and $G'$ reduces to the comparison of the respective system matrices $U$ and $U'$.
While conceptually simple, this quickly amounts to a non-trivial task due to the fact that the involved matrices grow exponentially with respect to the number of qubits.
Equivalence checking of quantum circuits has even been shown to be QMA-complete~\cite{janzingNonidentityCheckQMAcomplete2005}.
Nevertheless, several methods for this problem have been proposed~\cite{yamashitaFastEquivalencecheckingQuantum2010, burgholzerAdvancedEquivalenceChecking2021, viamontesCheckingEquivalenceQuantum2007,niemannEquivalenceCheckingMultilevel2014,wangXQDDbasedVerificationMethod2008,smithQuantumComputationalCompiler2019, amyLargescaleFunctionalVerification2019,hongTensorNetworkBased2020}.

\section{Dynamic Circuits \\and Resulting Problem}\label{sec:dyncirc}

The circuit model of quantum computing, as discussed in the previous section, has been the de-facto standard for designing quantum circuits to be executed on current generation quantum computers.
However, this describes quantum circuits in a \emph{static} fashion---with no opportunity to steer the computation in a direction based on outcomes of intermediate results.
Recently, IBM announced that their quantum computers now allow for interactions with classical computing instructions within the runtime of a quantum circuit---enabling what IBM refers to as \emph{dynamic quantum circuits}~\cite{ibmquantumQuantumCircuitsGet2021}.
In the following, we describe what constitutes these new kind of circuits and discuss the resulting challenges for checking the equivalence of quantum circuits that might contain dynamic circuit primitives.

\subsection{Dynamic Quantum Circuits \\and Their Benefits}

By allowing the interaction of real-time classical computations with the gates and measurements of traditional quantum circuits, the quantum circuit model reviewed in \autoref{sec:qc} is extended by non-unitary primitives such as \mbox{mid-circuit} measurements and resets as well as \mbox{classically-controlled} quantum operations.
As a consequence, circuits are no longer static, but rather \emph{dynamic}.

Eventually, these primitives will be necessary for quantum computers to achieve fault-tolerance by realizing quantum error correction schemes.
However, already in the near term, interesting use cases for teleportation~\cite{bennettTeleportingUnknownQuantum1993} and algorithms like \emph{Iterative  QPE} (IQPE,~\cite{dobsicekArbitraryAccuracyIterative2007}) arise that employ dynamic circuit primitives in order to, e.g., reduce the required number of qubits---a limited resource thus far.

For example, our running example, i.e., the QPE algorithm reviewed in \autoref{sec:qpe}, may exploit non-unitaries to reduce the number of qubits: 
Instead of an \mbox{$m$-qubit} register for computing the Fourier base representation of the unitary's phase, a single qubit is used and repeatedly measured.
Starting from the least significant bit of the resulting estimate \mbox{$\tilde\theta = 0.c_{m-1}\dots c_0$}, each measurement adds one bit of information to the estimated phase.
The result of each measurement then influences the rotation angles applied to the working qubit in the next iteration.
This requires the availability of the measurement results and application of quantum operations based on them within the coherence time of the quantum computer's qubits.
One of the first realizations of the IQPE algorithm on an actual system has recently been demonstrated by researchers from IBM Quantum on one of their devices~\cite{corcolesExploitingDynamicQuantum2021}.

\begin{figure}[t]
\centering
	\resizebox{0.8\linewidth}{!}{
\begin{tikzpicture}[label position=north west, every label/.style={inner sep=1pt}]
	\begin{yquant}[every post measurement control=direct, register/separation=1mm, operator/separation=0.5mm]
		qubit {$\ket{q_0} = \ket{0}$} q[1];
		qubit {$\ket{\psi} = \ket{1}$} q[+1];
		cbit {$c_\idx$} c[3];
		
		h q[0];
		[value=$\frac{12\pi}{8}$]
		phase q[1] | q[0];
		h q[0];
		measure q[0];
		cnot c[0] | q[0];
		discard q[0];

		init {$\ket{0}$} q[0];
		h q[0];
		[value=$\frac{6\pi}{8}$]
		phase q[1] | q[0];
		[value=$-\frac{\pi}{2}$]
		phase q[0] | c[0];
		h q[0];
		measure q[0];
		cnot c[1] | q[0];
		discard q[0];

		init {$\ket{0}$} q[0];
		h q[0];
		[value=$\frac{3\pi}{8}$]
		phase q[1] | q[0];
		[value=$-\frac{\pi}{4}$]
		phase q[0] | c[0];
		[value=$-\frac{\pi}{2}$]
		phase q[0] | c[1];
		h q[0];
		measure q[0];
		cnot c[2] | q[0];
		discard q[0];
	\end{yquant}
\end{tikzpicture}}\vspace*{-2mm}
\caption{Dynamic version of the QPE circuit from \autoref{fig:qpe}}\vspace*{-1em}
\label{fig:iqpe}
\end{figure}
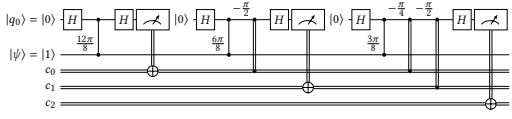

\begin{example}\label{ex:iqpe}
Assume again that, as in \autoref{ex:qpe}, we want to iteratively estimate the phase~$\theta$ of the unitary operator $U=p(\frac{3\pi}{8})$ corresponding to the eigenvector state $\ket{\psi}=\ket{1}$ up to a precision of three bits.
\autoref{fig:iqpe} shows an alternative quantum circuit utilizing dynamic circuit primitives. 
Instead of the $3$-qubit register considered before in \autoref{fig:qpe}, a single working qubit in combination with mid-circuit measurements, resets, and classically-controlled single-qubit rotations is used to iteratively compute individual bits of the phase estimate.
Compiling this circuit to an actual device requires no mapping at all, since only two qubits interact with each other.
As a consequence, the quantum cost of the resulting circuit is considerably reduced---significantly improving the expected fidelity when executing the circuit on an actual device.
\end{example}

\subsection{Resulting Problem}

Existing frameworks for verifying quantum circuits such as~\cite{yamashitaFastEquivalencecheckingQuantum2010, burgholzerAdvancedEquivalenceChecking2021, viamontesCheckingEquivalenceQuantum2007,niemannEquivalenceCheckingMultilevel2014,wangXQDDbasedVerificationMethod2008,smithQuantumComputationalCompiler2019, amyLargescaleFunctionalVerification2019, hongTensorNetworkBased2020} generally assume the circuit to only contain unitary operations.
Ultimately, only then it is possible to characterize the functionality of a quantum circuit as a unitary matrix.
With the availability of dynamic circuit primitives for conducting quantum computations, the question arises how circuits using these primitives can be verified.
After all, resets, measurements, and classically-controlled operations are all non-unitary operations.
As such, existing techniques cannot be applied in an out-of-the-box fashion.

Several theoretical works on quantum program and protocol verification exist that deal with dynamic quantum circuits, e.g.,~\cite{yingFloydHoareLogic2011, yingFoundationsQuantumProgramming2016}. However, their goal is to prove the correctness of an algorithm,~i.e., proving that it \enquote{works}, rather than to check the equivalence of two circuits. 
Recent works on the equivalence of dynamic quantum circuits based on quantum Mealy machines~\cite{wangEquivalenceCheckingSequential2021} and ensembles of linear operators~\cite{hongEquivalenceCheckingDynamic2021} show promise, but have only been evaluated on toy examples ($\approx 10$ qubits) and have not led to available software packages for equivalence checking yet.
In this work, we show that reinventing the wheel is not necessary to allow the usage of existing techniques and tools in combination with dynamic circuits.
To this end, we propose two different schemes targeted at two slightly different verification scenarios.

First, we consider the question whether two circuits $G$ and $G'$ which might contain non-unitaries are functionally equivalent as a whole---an important question when, e.g., evaluating alternative realizations of certain building blocks in large quantum algorithms.
Here, it has to be ensured that the alternative realization has the exact same functionality given \emph{any} input.
As already shown in \autoref{sec:compflowver}, given two circuits $G$ and $G'$ which only contain unitary operations, this reduces to the comparison between the corresponding unitary matrices $U$ and $U'$. 
We will show in \autoref{sec:circuit_transform} that any circuit containing non-unitary operations can be transformed to a circuit only containing unitary operations and no intermediate measurements by combining well known results from quantum information theory.
This way, all existing techniques for verifying the equivalence of two (static) quantum circuits are kept applicable for the broader class of dynamic circuits.

While the above technique conceptually allows to verify circuits containing non-unitaries, it requires to extend a circuits description by as many qubits as it contains mid-circuit resets.
Due to the exponential scaling of the resulting unitary functionality, the complexity of verifying such instances may prove too much to handle for existing tools. The following observation helps to derive an alternative for these cases:
In most quantum algorithms, the initial state of the computation can be assumed to be a fixed state (e.g., $\ket{0\dots 0}$).
Hence, it might not be necessary at all to ensure that two circuits are \emph{fully} functionally-equivalent, but rather that they produce the same distribution of measurement outcomes for the fixed input state, i.e., that they behave the same when executed on a quantum computer.
In \autoref{sec:density_sim}, we show that the probability distribution of a circuit containing non-unitaries can be iteratively extracted from classically simulating the circuit using any available classical quantum circuit simulator.

\section{Unitary Reconstruction\\ through Circuit Transformation}\label{sec:circuit_transform}

Dynamic circuit primitives allow to re-use qubits over the course of a quantum computation and to influence the execution based on classical measurement outcomes.
In order to employ existing verification tools for verifying circuits using these primitives, the circuit descriptions $G$ and $G'$ have to be transformed to facilitate comparisons of the form $U =^? U'$.
This is accomplished by transforming the dynamic circuit primitives to unveil the underlying unitary functionality.

Reset operations pose the first hurdle to overcome in this endeavour. 
Algorithmically, a reset can be interpreted as measuring a qubit, applying an $X$ operation conditioned on the measurement result being $\ket{1}$ and, then, discarding the measurement result.
Theoretically, any reset operation can be replaced by introducing a new qubit and applying all subsequent operations involving the qubit to be reset to the new qubit.
In this fashion, any \mbox{$n$-qubit} circuit containing $r$ reset instructions can be transformed to a circuit acting on $n+r$ qubits containing no reset primitives.

\begin{figure}[t]
	\centering
	\begin{tikzpicture}
		\node (qubits) {
		\resizebox{0.75\linewidth}{!}{
		\begin{tikzpicture}[label position=north west, every label/.style={inner sep=1pt}]
		\begin{yquant}[every post measurement control=direct, register/separation=1mm, operator/separation=0.5mm]
		qubit {$\ket{q_2} = \ket{0}$} q;
		qubit {$\ket{q_1} = \ket{0}$} q[+1];
		qubit {$\ket{q_0} = \ket{0}$} q[+1];
		qubit {$\ket{\psi} = \ket{1}$} q[+1];
		cbit {$c_\idx$} c[3];
		
		h q[2];
		[value=$\frac{12\pi}{8}$]
		phase q[3] | q[2];
		h q[2];
		measure q[2];
		cnot c[0] | q[2];
		discard q[2];

		h q[1];
		[value=$\frac{6\pi}{8}$]
		phase q[3] | q[1];
		[value=$-\frac{\pi}{2}$]
		phase q[1] | c[0];
		h q[1];
		measure q[1];
		cnot c[1] | q[1];
		discard q[1];

		h q[0];
		[value=$\frac{3\pi}{8}$]
		phase q[3] | q[0];
		[value=$-\frac{\pi}{4}$]
		phase q[0] | c[0];
		[value=$-\frac{\pi}{2}$]
		phase q[0] | c[1];
		h q[0];
		measure q[0];
		cnot c[2] | q[0];
		discard q[0];
		\end{yquant}
		\end{tikzpicture}}};
		\node [below= 1ex of qubits.south,inner sep=0pt]{\parbox{\linewidth}{\subcaption{Circuit after substituting new qubits for every reset\label{fig:reconstructionReset}}}};

		\node[below=1.8em of qubits] (delay) {
		\resizebox{0.7\linewidth}{!}{
		\begin{tikzpicture}[label position=north west, every label/.style={inner sep=1pt}]
		\begin{yquant}[register/separation=1mm, operator/separation=0.5mm]
		qubit {$\ket{q_2} = \ket{0}$} q;
		qubit {$\ket{q_1} = \ket{0}$} q[+1];
		qubit {$\ket{q_0} = \ket{0}$} q[+1];
		qubit {$\ket{\psi} = \ket{1}$} q[+1];
		
		h q[2];
		h q[1];
		h q[0];
		[value=$\frac{12\pi}{8}$]
		phase q[3] | q[2];
		[value=$\frac{6\pi}{8}$]
		phase q[3] | q[1];
		[value=$\frac{3\pi}{8}$]
		phase q[3] | q[0];
		
		h q[2];

		[value=$-\frac{\pi}{2}$]
		phase q[1] | q[2];
		h q[1];

		[value=$-\frac{\pi}{4}$]
		phase q[0] | q[2];
		[value=$-\frac{\pi}{2}$]
		phase q[0] | q[1];
		h q[0];
		align q[0-2];
		measure q[2];
		measure q[1];
		measure q[0];
		output {$c_0$} q[2];
		output {$c_1$} q[1];
		output {$c_2$} q[0];
		\end{yquant}
		\end{tikzpicture}}};
		\node [below= 1ex of delay.south,inner sep=0pt]{\parbox{\linewidth}{\subcaption{Circuit after applying deferred measurement principle\label{fig:reconstructionDelayed}}}};

	\end{tikzpicture}
	\vspace*{-5mm}
	\caption{Unitary reconstruction for IQPE circuit from \autoref{fig:iqpe}}
	\label{fig:reconstruction}
	\vspace*{-3mm}
\end{figure}
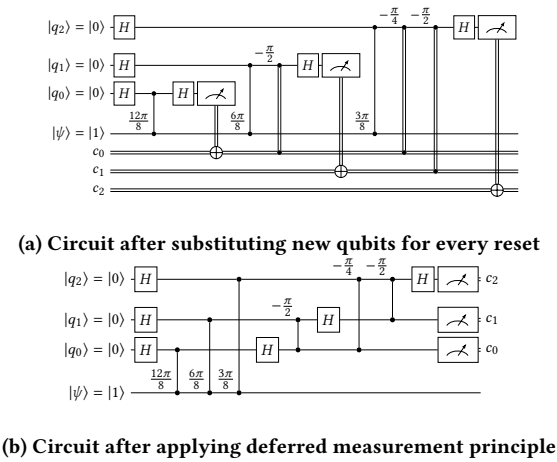

\begin{example}\label{ex:reset-reconstruction}
	Consider again the circuit for the $3$-bit precision IQPE algorithm from \autoref{ex:iqpe} shown in \autoref{fig:iqpe}. By iteratively replacing each of the reset operations with a new qubit and translating all subsequent gates to the newly introduced qubits, a circuit acting on four qubits results, as shown in \autoref{fig:reconstructionReset}.
\end{example}

Once qubit re-use is eliminated from a dynamic circuit, the only potentially non-unitary primitives remaining are mid-circuit measurements and classically-controlled operations conditioned on their result.
In order to get rid of these operations, we resort to one of the most fundamental results in quantum computing: the \emph{deferred measurement principle}~\cite{nielsenQuantumComputationQuantum2010}.
This principle states that delaying measurements until the end of a quantum computation does not affect the probability distribution of outcomes.
As a consequence, it follows that measurement and classical-conditioning on its result commute.
Thus, any mid-circuit measurement can be delayed until the very end of the quantum circuit---replacing any classically-controlled operations along the way by proper quantum operations controlled by the respective qubit.

\begin{example}\label{ex:measure-delay}
	Assume that all reset operations of the IQPE circuit from \autoref{ex:iqpe} have been eliminated, e.g., by transforming the circuit as described in \autoref{ex:reset-reconstruction}.
	Then, applying the deferred measurement principle in order to delay all measurements to the end of the circuit and replacing the phase rotations controlled by the measurement outcomes with phase gates controlled on the respective qubits, results in a circuit as shown in \autoref{fig:reconstructionDelayed}---free of dynamic circuit primitives.
\end{example}

By combining both aforementioned steps, i.e., substituting reset operations with \enquote{fresh} qubits and applying the deferred measurement principle, any dynamic quantum circuit (including non-unitaries) can be transformed to a representation composed of unitary descriptions only.
For one, this allows to verify that a dynamic circuit actually realizes the intended functionality of its static counterpart.

\begin{example}\label{ex:dynamic-eq}
Compare the transformed circuit obtained in \autoref{ex:measure-delay} (shown in \autoref{fig:reconstructionDelayed}) to the original QPE algorithm shown in \autoref{fig:qpe}.
Due to them actually being the same, it is easy to conclude that both circuits are indeed equivalent.
\end{example}

Note that it might seem that the proposed approach merely reverses the circuit construction or compilation process.
As witnessed in \autoref{ex:dynamic-eq}, there is a one-to-one relation between the transformed version of the IQPE circuit shown in \autoref{fig:reconstructionDelayed} and the original QPE circuit shown in \autoref{fig:qpe}.
As such, it could be argued that there is nothing to be gained from using the technique. 
However, this is not the case, as almost no assumptions are made about the relation between $G$ and $G'$ in general.
Indeed, the only requirement is that the transformed versions of both circuits have the same number of primary inputs and outputs.
The proposed transformation scheme \enquote{touches} nothing but reset, measurement, and \mbox{classically-controlled} operations---which are ``reversed''.

Conceptually, this approach allows to verify circuits containing non-unitaries, at the cost of extending a circuit's description by as many qubits as it contains mid-circuit resets.
Since the resulting unitary functionality scales exponentially with the number of qubits, the complexity of verifying such instances increases quickly.
However, this is an inevitable increase whenever verifying whether a dynamic implementation (acting on $n_{\mathit{dyn}}$ qubits and using $r$ resets) still realizes the same functionally as a static counterpart (acting on $n_{\mathit{static}}$ qubits).
Since in that case, $n_{\mathit{dyn}} + r = n_{\mathit{static}}$, the proposed scheme augments the dynamic circuit just enough to facilitate comparisons of the form $U =^? U'$.

\section{Extracting the Measurement Outcome Distribution by Simulation}\label{sec:density_sim}

Although verification methodologies such as~\cite{burgholzerAdvancedEquivalenceChecking2021, yamashitaFastEquivalencecheckingQuantum2010} frequently allow to reduce the complexity of the verification by exploiting the reversibility of quantum operations, they might not be able to handle this immense complexity in the worst case.
Motivated by the fact that most high-level quantum algorithms assume a fixed input state, we argue that it might be sufficient to show that two realizations of such an algorithm produce the same distribution of measurement probabilities given the fixed input state.
Verifying that two circuits $G$ and $G'$, which only contain unitary operations, produce equivalent probability distributions given a particular input state~$\ket{\psi}$ amounts to classically simulating both computations with~$\ket{\psi}$ as input and computing the overlap between the measurement probabilities described by the resulting state vectors.

However, in the presence of dynamic circuit primitives, the concept of a state vector responsible for producing the circuit's measurement outcome distribution (e.g., the probabilities of the individual bitstrings in the IQPE algorithm) does no longer make sense.
This is due to the non-unitary nature of the dynamic circuit primitives, that no longer allow to deterministically simulate the quantum circuit in one go using quantum circuit simulators such as~\cite{zulehnerAdvancedSimulationQuantum2019,guerreschiIntelQuantumSimulator2020,villalongaFlexibleHighperformanceSimulator2019}.
For example, each time a reset operation is encountered this would technically require the calculation of the partial trace of the system over the particular qubit (and reinitializing it to $\ket{0}$). However, the partial trace is an operation that maps pure states to mixed states.
One possible approach for solving this problem would be to repeatedly simulate the dynamic circuit and stochastically realize dynamic circuit primitives such as measurements and resets.
However, one would have to perform huge amounts of individual runs in order to reason about the output distribution in a statistically significant way.
Another approach requires leaving the pure state picture and using a density matrix simulator (such as, e.g.,~\cite{viamontesGraphbasedSimulationQuantum2004,grurlConsideringDecoherenceErrors2020,liDensityMatrixQuantum2020}).
Although these simulators can naturally handle resets, \mbox{mid-circuit} measurements, and \mbox{classic-controlled} operations, they also do not allow to determine the complete distribution of (intermediate) measurement outcomes via a single simulation run, but only the density matrix for a particular set of measurements.

In the following, we propose a technique that allows to extract the complete set of measurement probabilities for a dynamic circuit given a particular input state.
To this end, consider a quantum circuit $G$ involving $m$ measurements. 
Then, each measurement during the circuit simulation constitutes a branching point where the probabilities of the qubit to be measured are check-pointed and the simulation splits into two independent simulations: one assuming the measurement outcome is $\ket{0}$ and the other one assuming the outcome is~$\ket{1}$.
Depending on the outcome being $\ket{0}$ or $\ket{1}$, a subsequent reset operation is translated to a no-op or an $X$ gate, while any classically-controlled operation is ignored or applied, respectively.
The probability of observing a particular basis state $\ket{i} = \ket{(i_{m-1}\dots i_0)_2}$ can then be reconstructed from the product of the check-pointed probabilities along the path of simulations corresponding to the outcomes $i_{0}$ to $i_{m-1}$. 

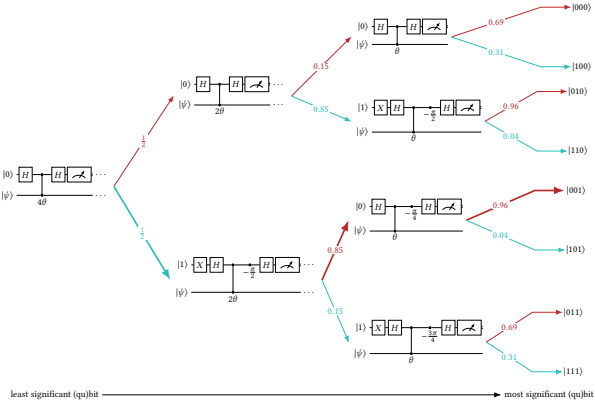
\begin{figure}[t]
	\centering
	\resizebox{0.95\linewidth}{!}{
	\begin{tikzpicture}
		\node (c0) {
		\begin{tikzpicture}[label position=south, every label/.style={inner sep=1pt}]
		\begin{yquant}[register/separation=2mm, operator/separation=0.5mm]
		qubit {$\ket{0}$} q;
		qubit {$\ket{\psi}$} q[+1];
		h q[0];
		[value=$4\theta$]
		phase q[1] | q[0];
		h q[0];
		measure q[0];
		align q;
		[draw=none]
		box {$\dots$} q;
		discard q;
		\end{yquant}
		\end{tikzpicture}};
		
		\node[above right=1.5cm and 2cm of c0] (c1_0) {
		\begin{tikzpicture}[label position=south, every label/.style={inner sep=1pt}]
		\begin{yquant}[register/separation=2mm, operator/separation=0.5mm]
		qubit {$\ket{0}$} q;
		qubit {$\ket{\psi}$} q[+1];
		h q[0];
		[value=$2\theta$]
		phase q[1] | q[0];
		h q[0];
		measure q[0];
		align q;
		[draw=none]
		box {$\dots$} q;
		discard q;
		\end{yquant}
		\end{tikzpicture}};
		
		\node[below right=1.5cm and 1.9cm of c0] (c1_1) {
		\begin{tikzpicture}[label position=south, every label/.style={inner sep=1pt}]
		\begin{yquant}[register/separation=2mm, operator/separation=0.5mm]
		qubit {$\ket{1}$} q;
		qubit {$\ket{\psi}$} q[+1];
		x q[0];
		h q[0];
		[value=$2\theta$]
		phase q[1] | q[0];
		[value=$-\frac{\pi}{2}$]
		phase q[0];
		h q[0];
		measure q[0];
		align q;
		[draw=none]
		box {$\dots$} q;
		discard q;
		\end{yquant}
		\end{tikzpicture}};
		
		\node[above right=0.5cm and 2cm of c1_0] (c2_0) {
		\begin{tikzpicture}[label position=south, every label/.style={inner sep=1pt}]
		\begin{yquant}[register/separation=2mm, operator/separation=0.5mm]
		qubit {$\ket{0}$} q;
		qubit {$\ket{\psi}$} q[+1];
		h q[0];
		[value=$\theta$]
		phase q[1] | q[0];
		h q[0];
		measure q[0];
		\end{yquant}
		\end{tikzpicture}};
	
		\node[below right=-0.75cm and 2cm of c1_0] (c2_1) {
		\begin{tikzpicture}[label position=south, every label/.style={inner sep=1pt}]
		\begin{yquant}[register/separation=2mm, operator/separation=0.5mm]
		qubit {$\ket{1}$} q;
		qubit {$\ket{\psi}$} q[+1];
		x q[0];
		h q[0];
		[value=$\theta$]
		phase q[1] | q[0];
		[value=$-\frac{\pi}{2}$]
		phase q[0];
		h q[0];
		measure q[0];
		\end{yquant}
		\end{tikzpicture}};
		
		\node[above right=0.3cm and 0.9cm of c1_1] (c2_2) {
		\begin{tikzpicture}[label position=south, every label/.style={inner sep=1pt}]
		\begin{yquant}[register/separation=2mm, operator/separation=0.5mm]
		qubit {$\ket{0}$} q;
		qubit {$\ket{\psi}$} q[+1];
		h q[0];
		[value=$\theta$]
		phase q[1] | q[0];
		[value=$-\frac{\pi}{4}$]
		phase q[0];
		h q[0];
		measure q[0];
		\end{yquant}
		\end{tikzpicture}};
		
		\node[below right=0.35cm and 0.9cm of c1_1] (c2_3) {
		\begin{tikzpicture}[label position=south, every label/.style={inner sep=1pt}]
		\begin{yquant}[register/separation=2mm, operator/separation=0.5mm]
		qubit {$\ket{1}$} q;
		qubit {$\ket{\psi}$} q[+1];
		x q[0];
		h q[0];
		[value=$\theta$]
		phase q[1] | q[0];
		[value=$-\frac{3\pi}{4}$]
		phase q[0];
		h q[0];
		measure q[0];
		\end{yquant}
		\end{tikzpicture}};
		
		\draw[-Latex,semithick, inner sep=1pt,RedEdge] (c0.east) -- (c1_0.west) node [midway,fill=white] {$\frac{1}{2}$};
		\draw[-Latex,ultra thick, inner sep=1pt,BlueEdge] (c0.east) -- (c1_1.west) node [midway,fill=white] {$\frac{1}{2}$};
		
		\draw[-Latex,semithick, inner sep=1pt,RedEdge] (c1_0.east) -- (c2_0.west) node [midway,fill=white] {$0.15$};
		\draw[-Latex,semithick, inner sep=1pt,BlueEdge] (c1_0.east) -- (c2_1.west) node [midway,fill=white] {$0.85$};
		
		\draw[-Latex,ultra thick, inner sep=1pt,RedEdge] (c1_1.east) -- (c2_2.west) node [midway,fill=white] {$0.85$};
		\draw[-Latex,semithick, inner sep=1pt,BlueEdge] (c1_1.east) -- (c2_3.west) node [midway,fill=white] {$0.15$};
		
		\draw[-Latex,semithick, inner sep=1pt,RedEdge] (c2_0.east) -- ++(3,1) node [midway,fill=white] {$0.69$} -- ++(1,0) node[black,right] {$\ket{000}$};
		\draw[-Latex,semithick, inner sep=1pt,BlueEdge] (c2_0.east) -- ++(3,-1) node [midway,fill=white] {$0.31$} -- ++(1,0) node[black,right] {$\ket{100}$};
		
		\draw[-Latex,semithick, inner sep=1pt,RedEdge] (c2_1.east) -- ++(1.75,1) node [midway,fill=white] {$0.96$} -- ++(1,0) node[black,right] {$\ket{010}$};
		\draw[-Latex,semithick, inner sep=1pt,BlueEdge] (c2_1.east) -- ++(1.75,-1) node [midway,fill=white] {$0.04$} -- ++(1,0) node[black,right] {$\ket{110}$};
		
		\draw[-Latex,ultra thick, inner sep=1pt,RedEdge] (c2_2.east) -- ++(2.3,1) node [midway,fill=white] {$0.96$} -- ++(1,0) node[black,right] {$\ket{001}$};
		\draw[-Latex,semithick, inner sep=1pt,BlueEdge] (c2_2.east) -- ++(2.3,-1) node [midway,fill=white] {$0.04$} -- ++(1,0) node[black,right] {$\ket{101}$};
		
		\draw[-Latex,semithick, inner sep=1pt,RedEdge] (c2_3.east) -- ++(1.55,1) node [midway,fill=white] {$0.69$} -- ++(1,0) node[black,right] {$\ket{011}$};
		\draw[-Latex,semithick, inner sep=1pt,BlueEdge] (c2_3.east) -- ++(1.55,-1) node [midway,fill=white] {$0.31$} -- ++(1,0) node[black,right] {$\ket{111}$};
		
		\node (ls) at ($(c0)+(0,-7)$) {least significant (qu)bit};
		\node (ms) at ($(ls.east)+(15.0,0)$) {most significant (qu)bit};
		\draw[thick, -Latex] (ls) -- (ms);
	\end{tikzpicture}
	}
	\caption{Measurement outcome distribution extraction for the IQPE circuit with $\theta=\frac{3\pi}{8}$. \textcolor{RedEdge}{Red} and \textcolor{BlueEdge}{blue} arrows denote the \mbox{$\ket{0}$-} and \mbox{$\ket{1}$-successor}, respectively.}\label{fig:state-extraction}
\end{figure}

\begin{example}\label{ex:extraction}
	Consider again the IQPE algorithm for estimating the phase~$\theta$ of $U=p(\frac{3\pi}{8})$ corresponding to the eigenstate $\ket{\psi}=\ket{1}$ up to a precision of three bits, as shown in \autoref{fig:iqpe}. 
	The circuit contains a total of $m=3$ measurements (necessary for the $3$-bit precisison) and uses the fixed input state \mbox{$\ket{000}\otimes\ket{\psi}=\ket{0001}$}.
	Iteratively simulating the circuit, check-pointing the probabilities at each of the measurements, and adjusting the subsequent circuit parts to be simulated accordingly, results in a computational flow as illustrated in \autoref{fig:state-extraction}.
	There, \textcolor{RedEdge}{red} arrows denote the \mbox{$\ket{0}$-successor}, while \textcolor{BlueEdge}{blue} arrows denote the \mbox{$\ket{1}$-successor},~i.e., the subsequent computations upon measuring $\ket{0}$ or $\ket{1}$, respectively.
	The path indicated in bold represents the extraction of the probability for the $\ket{001}$ basis state---resulting in $\frac{1}{2} * 0.85 * 0.96 \approx 0.408$. 
\end{example}

Extracting the distribution of measurement outcomes of a dynamic circuit in this fashion naturally requires a total of $2^m$ individual simulations, where $m$ is the number of \mbox{mid-circuit} measurements.
However, large parts of the simulations can be shared in between simulation runs.
For example, the circuit up until the first checkpoint only needs to be simulated once, while two simulations are necessary up until the second checkpoint, and so on.
In general, the $k^\mathit{th}$ \mbox{sub-circuit} needs to be simulated in at most $2^k$ variations.
If any measurement along a path produces a probability of zero, further simulations along that path need not be started at all.
In addition, the individual simulations in between checkpoints are completely independent from another and, hence, are embarrassingly parallelizable.
On top of that, each of these \mbox{sub-circuits} consists of a much smaller number of gates and acts on far fewer qubits than the whole dynamic circuit's static counterpart.
As a consequence, the complete measurement outcome distribution can be efficiently extracted in many cases, even though exponentially many simulations might be required in the worst case.

\section{Experimental Evaluation}\label{sec:results}

The methods proposed above can be implemented on top of any existing verification or simulation tool, respectively.
In order to demonstrate that the proposed schemes indeed allow to efficiently handle non-unitaries in equivalence checking flows, we exemplarily implemented them on top of the \mbox{open-source} quantum circuit equivalence checking tool \emph{QCEC}~\cite{burgholzerQCEC} that is publicly available at \mbox{\url{https://github.com/iic-jku/qcec}}.
The tool supports complete functional verification (as considered in \autoref{sec:circuit_transform}) as well as simulative verification (as considered in \autoref{sec:density_sim}).

As benchmarks we consider various instances of the famous Bernstein-Vazirani algorithm~\cite{bernsteinQuantumComplexityTheory1997}, the Quantum Fourier Transform, and the QPE algorithm, which was used as running example throughout this work.
For each static algorithm, a dynamic realization has been derived~\cite{nationHowMeasureReset2021, griffithsSemiclassicalFourierTransform1996, dobsicekArbitraryAccuracyIterative2007}.
These algorithms are a good fit for evaluating the overhead of the proposed schemes, as they feature all the hurdles of dynamic quantum circuits that have to be overcome for verifying their equivalence.
All evaluations have been conducted on a machine equipped with an \mbox{AMD Ryzen 9 5950X} CPU and \SI{64}{\gibi\byte} RAM running \mbox{Ubuntu 20.04}.
\autoref{tab:eval} summarizes the obtained results. 
To this end, it first lists the number of qubits $n$ as well as the number of gates $\vert G \vert$ of the original (static) and the dynamic circuit, respectively. 
Then, the runtime $t_\mathit{trans}$ of the transformation scheme (as proposed in \autoref{sec:circuit_transform}) is listed along the time $t_\mathit{ver}$ it took to verify the equivalence of both circuits.
Finally, $t_\mathit{extract}$ denotes the runtime of the extraction scheme (proposed in \autoref{sec:density_sim}) applied to the dynamic circuit, while $t_\mathit{sim}$ denotes the runtime of the classical simulation of the original (static) circuit.

\begin{table*}
	\centering
	\caption{Experimental Evaluations}\label{tab:eval}\vspace*{-3mm}
	\footnotesize
	\resizebox{0.53\linewidth}{!}{
		\begin{tabular}{rrrrrrrr}\toprule
			\multicolumn{2}{c}{Static} & \multicolumn{2}{c}{Dynamic} & \multicolumn{2}{c}{Full Functional Verification} & \multicolumn{2}{c}{Fixed Input State} \\
			\cmidrule(r){1-2}\cmidrule(lr){3-4}\cmidrule(lr){5-6}\cmidrule(lr){7-8}
			{$n$} & {$\vert G \vert$} & $n$ & {$\vert G\vert$} & {$t_{trans}\,[\si{\second}]$} & {$t_{ver}\,[\si{\second}]$} & {$t_\mathit{extract}\,[\si{\second}]$} & {$t_\mathit{sim}\,[\si{\second}]$} \\\midrule\\
			\multicolumn{8}{c}{Bernstein-Vazirani} \\\bottomrule
			\begin{filecontents*}{results_bv.csv}
2,3,2,4,1.37E-06,1.48E-05,3.97E-06,8.11E-06
3,6,2,9,6.70E-06,3.28E-05,5.74E-06,9.49E-06
4,8,2,13,5.50E-06,4.71E-05,7.32E-06,1.26E-05
5,10,2,17,3.27E-06,3.42E-05,8.02E-06,1.78E-05
6,13,2,22,3.35E-06,6.00E-05,1.01E-05,2.68E-05
7,16,2,27,4.21E-06,8.59E-05,1.14E-05,3.36E-05
8,19,2,32,4.91E-06,0.000115137,1.34E-05,4.34E-05
9,22,2,37,5.39E-06,0.000147296,1.58E-05,5.54E-05
10,24,2,41,6.40E-06,0.000170856,1.66E-05,6.59E-05
11,27,2,46,7.22E-06,0.000219385,1.87E-05,8.09E-05
12,30,2,51,8.11E-06,0.000268534,2.01E-05,9.49E-05
13,32,2,55,8.49E-06,0.000309172,2.18E-05,0.000110027
14,35,2,60,9.48E-06,0.000345681,2.40E-05,0.000129686
15,37,2,64,1.05E-05,0.00039372,2.67E-05,0.000159506
16,39,2,68,1.15E-05,0.00043646,2.82E-05,0.000166246
17,41,2,72,1.29E-05,0.000483098,2.97E-05,0.000187996
18,44,2,77,1.34E-05,0.000596565,3.25E-05,0.000209605
19,47,2,82,1.47E-05,0.000733712,3.49E-05,0.000235344
20,49,2,86,1.58E-05,0.000716412,3.73E-05,0.000286363
21,52,2,91,1.67E-05,0.00078973,3.88E-05,0.000284313
22,55,2,96,1.76E-05,0.000862289,4.19E-05,0.000320032
23,58,2,101,1.89E-05,0.000963106,4.47E-05,0.000350981
24,60,2,105,2.02E-05,0.00104131,4.75E-05,0.000396721
25,63,2,110,2.14E-05,0.00121128,5.05E-05,0.00041573
26,66,2,115,2.43E-05,0.0012576,5.10E-05,0.000446969
27,69,2,120,2.55E-05,0.00139698,5.36E-05,0.000486579
28,71,2,124,2.70E-05,0.00152008,5.71E-05,0.000522798
29,73,2,128,2.85E-05,0.00153443,5.77E-05,0.000555866
30,75,2,132,2.90E-05,0.00161397,6.05E-05,0.000616315
31,78,2,137,2.95E-05,0.00172305,6.18E-05,0.000632945
32,81,2,142,3.13E-05,0.00186979,6.49E-05,0.000684484
33,84,2,147,3.38E-05,0.00201734,6.81E-05,0.000731632
34,86,2,151,3.79E-05,0.00224653,7.32E-05,0.000799291
35,88,2,155,3.98E-05,0.00226175,7.28E-05,0.00081666
36,90,2,159,4.07E-05,0.00340932,7.52E-05,0.000875558
37,93,2,164,4.44E-05,0.0026972,7.82E-05,0.000915568
38,96,2,169,4.50E-05,0.00282884,7.94E-05,0.000966667
39,98,2,173,4.65E-05,0.00294718,8.60E-05,0.00104581
40,101,2,178,4.86E-05,0.00314341,8.65E-05,0.00113615
41,104,2,183,4.92E-05,0.00350496,8.74E-05,0.00116031
42,106,2,187,5.27E-05,0.00346177,9.09E-05,0.00122798
43,108,2,191,5.47E-05,0.00353598,9.28E-05,0.00130062
44,111,2,196,5.63E-05,0.00372253,9.54E-05,0.00136115
45,113,2,200,5.39E-05,0.00366775,9.72E-05,0.00142157
46,116,2,205,5.60E-05,0.00388877,9.96E-05,0.00150761
47,119,2,210,5.92E-05,0.00424322,0.000102417,0.00155618
48,121,2,214,6.03E-05,0.00433401,0.000104217,0.00160986
49,123,2,218,6.81E-05,0.00453706,0.000106508,0.0016654
50,126,2,223,9.54E-05,0.00468722,0.000109557,0.00167535
51,129,2,228,6.74E-05,0.00478188,0.000114587,0.00184674
52,132,2,233,7.04E-05,0.00517536,0.000113888,0.00193453
53,135,2,238,7.01E-05,0.00546742,0.000120397,0.00212612
54,138,2,243,7.23E-05,0.00555777,0.000120987,0.00215235
55,140,2,247,7.44E-05,0.00571829,0.000122577,0.00221336
56,142,2,251,7.70E-05,0.00594333,0.000124647,0.00229183
57,144,2,255,7.83E-05,0.00597321,0.000127217,0.00234259
58,146,2,259,8.22E-05,0.00605969,0.000129636,0.00244352
59,148,2,263,8.38E-05,0.0064556,0.000132266,0.00252938
60,151,2,268,8.81E-05,0.00668045,0.000133477,0.00256823
61,154,2,273,8.93E-05,0.00807174,0.000137717,0.00267229
62,157,2,278,0.000108137,0.00865608,0.000141287,0.00281973
63,159,2,282,0.000126097,0.00929427,0.000143077,0.00294254
64,162,2,287,0.000118947,0.00910588,0.000141616,0.00292281
65,164,2,291,0.000114347,0.00895289,0.000145696,0.0030079
66,167,2,296,0.000117827,0.0094577,0.000148997,0.00320708
67,170,2,301,0.000111537,0.00937464,0.000153527,0.00320828
68,172,2,305,0.000115347,0.00957911,0.000153316,0.00333244
69,175,2,310,0.000116527,0.00979309,0.000152656,0.00345455
70,178,2,315,0.000120317,0.00985707,0.000155336,0.00369075
71,181,2,320,0.000119977,0.0102449,0.000153916,0.00363997
72,183,2,324,0.000122317,0.0107948,0.000164486,0.00378374
73,186,2,329,0.000129616,0.0106485,0.000160766,0.00384175
74,188,2,333,0.000128227,0.0106007,0.000171136,0.00393884
75,191,2,338,0.000128107,0.0111769,0.000168926,0.00405142
76,193,2,342,0.000132577,0.0113414,0.000172685,0.00423133
77,196,2,347,0.000143616,0.012444,0.000198465,0.00436726
78,198,2,351,0.000154216,0.0126701,0.000173336,0.00437252
79,200,2,355,0.000158646,0.0128962,0.000179426,0.00448871
80,203,2,360,0.000152066,0.0130606,0.000183106,0.0046629
81,206,2,365,0.000152116,0.0130714,0.000183976,0.00470472
82,209,2,370,0.000167686,0.0143198,0.000188416,0.00499628
83,211,2,374,0.000170325,0.0140307,0.000190075,0.00495894
84,213,2,378,0.000176486,0.0142031,0.000192125,0.00510791
85,216,2,383,0.000180406,0.0147267,0.000192915,0.0051693
86,219,2,388,0.000180136,0.0143823,0.000197885,0.00530616
87,221,2,392,0.000165136,0.0145883,0.000200855,0.00567834
88,224,2,397,0.000169026,0.0151254,0.000206035,0.00576733
89,226,2,401,0.000170606,0.0150773,0.000215084,0.00590101
90,228,2,405,0.000195816,0.0153906,0.000209244,0.005969
91,231,2,410,0.000178936,0.0155823,0.000214254,0.00614587
92,233,2,414,0.000182296,0.0160831,0.000218395,0.00640293
93,235,2,418,0.000183615,0.016226,0.000218565,0.00641101
94,237,2,422,0.000188295,0.0165012,0.000222534,0.0064704
95,240,2,427,0.000195826,0.0173844,0.000241254,0.00671447
96,242,2,431,0.000195555,0.0174398,0.000229554,0.00696806
97,244,2,435,0.000198906,0.0175719,0.000231235,0.00687469
98,247,2,440,0.000204685,0.0181218,0.000235835,0.00694981
99,249,2,444,0.000207285,0.0185919,0.000237074,0.00707612
100,251,2,448,0.000211295,0.0187138,0.000243934,0.00740521
101,254,2,453,0.000221295,0.0192276,0.000247714,0.00765297
102,257,2,458,0.000225505,0.0197635,0.000246484,0.00757845
103,260,2,463,0.000226185,0.0202785,0.000252214,0.00779995
104,263,2,468,0.000232294,0.0206223,0.000254874,0.00782729
105,265,2,472,0.000234084,0.0209199,0.000262434,0.00798477
106,267,2,476,0.000238604,0.0209669,0.000263053,0.00847387
107,270,2,481,0.000242844,0.0217248,0.000284663,0.00857617
108,273,2,486,0.000247814,0.0224343,0.000271934,0.00849814
109,275,2,490,0.000251634,0.0228103,0.000274483,0.00859715
110,277,2,494,0.000255564,0.0231729,0.000276183,0.00904312
111,279,2,498,0.000267133,0.0233848,0.000278173,0.00894364
112,281,2,502,0.000278913,0.0234192,0.000281143,0.00910542
113,283,2,506,0.000265903,0.0236387,0.000286703,0.00943198
114,285,2,510,0.000272223,0.0238479,0.000290913,0.00984317
115,287,2,514,0.000273873,0.0247846,0.000292953,0.00989136
116,289,2,518,0.000277963,0.0250451,0.000296392,0.0101313
117,291,2,522,0.000281603,0.0259991,0.000298682,0.010405
118,293,2,526,0.000284233,0.0258499,0.000301333,0.0103816
119,295,2,530,0.000287183,0.0263528,0.000311733,0.0111507
120,297,2,534,0.000291423,0.0267073,0.000309132,0.0108709
121,300,2,539,0.000300293,0.0276468,0.000310422,0.010742
122,303,2,544,0.000303843,0.0278117,0.000327622,0.0108628
123,305,2,548,0.000329592,0.0263704,0.000319912,0.0111323
124,307,2,552,0.000316072,0.0280735,0.000327312,0.0119319
125,310,2,557,0.000317992,0.0269229,0.000333422,0.0118399
126,312,2,561,0.000356172,0.0278108,0.000330562,0.0121005
127,314,2,565,0.000336632,0.0273855,0.000344811,0.0125118
128,317,2,570,0.000331632,0.0299686,0.000352232,0.0124263
\end{filecontents*}
\csvreader[range=120-127, no head, late after line=\\, late after last line=\\\bottomrule, separator=comma]{results_bv.csv}{}{%
				\tablenum[table-format=3.0, round-integer-to-decimal=false]{\csvcoli} & 
				\tablenum[table-format=4.0, round-integer-to-decimal=false]{\csvcolii} & 
				\tablenum[table-format=1.0, round-integer-to-decimal=false]{\csvcoliii} & 
				\tablenum[table-format=4.0, round-integer-to-decimal=false]{\csvcoliv} & 
				\tablenum[table-format=1.4, round-mode = places,round-precision = 4, scientific-notation=fixed, fixed-exponent=0]{\csvcolv} & 
				\tablenum[table-format=3.3, round-mode = places,round-precision = 3, group-minimum-digits = 4]{\csvcolvi} &
				\tablenum[table-format=3.4, round-mode = places,round-precision = 4, scientific-notation=fixed, fixed-exponent=0]{\csvcolvii} & 
				\tablenum[table-format=1.3, round-mode = places,round-precision = 3]{\csvcolviii}}\\
			\multicolumn{8}{c}{Quantum Fourier Transform} \\\bottomrule		
			\begin{filecontents*}{results_qft.csv}
1,1,1,2,8.10E-07,6.93E-06,4.22E-06,4.04E-06
2,3,1,6,3.87E-06,7.55E-06,6.56E-06,4.95E-06
3,6,1,11,3.31E-06,1.49E-05,8.90E-06,8.04E-06
4,10,1,17,4.64E-06,3.21E-05,1.75E-05,1.37E-05
5,15,1,24,6.03E-06,4.11E-05,3.53E-05,2.01E-05
6,21,1,32,8.13E-06,6.65E-05,7.12E-05,3.08E-05
7,28,1,41,1.03E-05,9.69E-05,0.000151847,4.44E-05
8,36,1,51,1.26E-05,0.000137376,0.000325223,6.39E-05
9,45,1,62,1.55E-05,0.000196055,0.000704603,8.53E-05
10,55,1,74,2.03E-05,0.000251434,0.00151937,0.000109597
11,66,1,87,2.44E-05,0.000330002,0.00320827,0.000143557
12,78,1,101,2.95E-05,0.00041666,0.00679309,0.000182225
13,91,1,116,3.53E-05,0.000533657,0.0144669,0.000227175
14,105,1,132,4.36E-05,0.000656094,0.0306646,0.000310693
15,120,1,149,5.33E-05,0.000796692,0.0650525,0.000362691
16,136,1,167,6.09E-05,0.000962537,0.137566,0.000415241
17,153,1,186,6.86E-05,0.00117872,0.290326,0.000472549
18,171,1,206,8.01E-05,0.00132628,0.610018,0.000589216
19,190,1,227,9.19E-05,0.00151923,1.31094,0.000677884
20,210,1,249,0.000105328,0.00178542,2.68059,0.00080666
21,231,1,272,0.000118727,0.00223342,5.64979,0.000923208
22,253,1,296,0.000134117,0.00236088,11.9683,0.00102821
23,276,1,321,0.000151687,0.00278224,24.8242,0.00109935
24,300,1,347,0.000173566,0.00322239,52.4281,0.00123912
25,325,1,374,0.000192226,0.0037249,107.316,0.00128319
26,351,1,402,0.000224895,0.00435268,223.419,0.00158881
27,378,1,431,0.000250514,0.00648179,,0.00228321
28,406,1,461,0.000271774,0.0058601,,0.00232063
29,435,1,492,0.000295823,0.00627496,,0.00212155
30,465,1,524,0.000567547,0.00731692,,0.00239725
31,496,1,557,0.000390951,0.0088027,,0.00253084
32,528,1,591,0.00041872,0.00973213,,0.00274703
33,561,1,626,0.000450699,0.0105483,,0.00300387
34,595,1,662,0.000462929,0.0113755,,0.00324525
35,630,1,699,0.000545907,0.0116415,,0.00416764
36,666,1,737,0.000601945,0.0140925,,0.00424339
37,703,1,776,0.000658405,0.0148527,,0.004362
38,741,1,816,0.000699914,0.0153118,,0.00456777
39,780,1,857,0.000773512,0.017807,,0.00489844
40,820,1,899,0.000855769,0.0185575,,0.00523571
41,861,1,942,0.000945737,0.0206044,,0.00584342
42,903,1,986,0.00101614,0.0229238,,0.0065589
43,946,1,1031,0.00109559,0.0263181,,0.00677059
44,990,1,1077,0.00118431,0.0301117,,0.00792461
45,1035,1,1124,0.00126694,0.0304836,,0.0095735
46,1081,1,1172,0.00140503,0.0273735,,0.00949902
47,1128,1,1221,0.0014023,0.0291698,,0.0103351
48,1176,1,1271,0.00153829,0.0302846,,0.0108282
49,1225,1,1322,0.00167295,0.0347246,,0.0114711
50,1275,1,1374,0.00181022,0.0390012,,0.012245
51,1326,1,1427,0.00190896,0.04234,,0.013165
52,1378,1,1481,0.00208164,0.0429556,,0.0141308
53,1431,1,1536,0.00225255,0.0493011,,0.0149448
54,1485,1,1592,0.00233679,0.0478495,,0.0156243
55,1540,1,1649,0.00263546,0.0480716,,0.0165266
56,1596,1,1707,0.00231194,0.0573415,,0.0174736
57,1653,1,1766,0.00261608,0.0599267,,0.0189081
58,1711,1,1826,0.00265827,0.0652581,,0.0192568
59,1770,1,1887,0.0029999,0.065742,,0.0221979
60,1829,1,1949,0.00303289,0.0655937,,0.0206498
61,1888,1,2012,0.00319153,0.063615,,0.0220108
62,1947,1,2076,0.00335044,0.0668173,,0.0228485
63,2006,1,2141,0.00350167,0.0761331,,0.023455
64,2065,1,2207,0.00373494,0.075714,,0.0254744
65,2124,1,2274,0.00392111,0.0853405,,0.0264429
66,2183,1,2342,0.00411292,0.0770764,,0.0268163
67,2242,1,2411,0.00435869,0.0904833,,0.0291854
68,2301,1,2481,0.00465598,0.0837886,,0.0255294
69,2360,1,2552,0.00483755,0.094952,,0.0265776
70,2419,1,2624,0.005213,0.0887371,,0.0300491
71,2478,1,2697,0.00541461,0.0889016,,0.0314922
72,2537,1,2771,0.00565233,0.100254,,0.0326383
73,2596,1,2846,0.00600432,0.0981167,,0.032426
74,2655,1,2922,0.0062649,0.102972,,0.0349776
75,2714,1,2999,0.00663549,0.11469,,0.0366751
76,2773,1,3077,0.00695051,0.116891,,0.0367564
77,2832,1,3156,0.00727556,0.114227,,0.0373454
78,2891,1,3236,0.00762159,0.124709,,0.0397591
79,2950,1,3317,0.00788686,0.127991,,0.0403423
80,3009,1,3399,0.0082119,0.130668,,0.0456189
81,3068,1,3482,0.00853635,0.128817,,0.0516378
82,3127,1,3566,0.00911623,0.139802,,0.0532676
83,3186,1,3651,0.00940238,0.139503,,0.052125
84,3245,1,3737,0.00978738,0.141431,,0.0554683
85,3304,1,3824,0.0102171,0.146306,,0.0554943
86,3363,1,3912,0.0107393,0.146191,,0.0559642
87,3422,1,4001,0.0110611,0.156021,,0.0590672
88,3481,1,4091,0.0116916,0.15193,,0.0601475
89,3540,1,4182,0.0120915,0.162875,,0.0615432
90,3599,1,4274,0.01278,0.175544,,0.0679768
91,3658,1,4367,0.0132161,0.173632,,0.0710062
92,3717,1,4461,0.0137594,0.178004,,0.0690474
93,3776,1,4556,0.0142989,0.179013,,0.0697927
94,3835,1,4652,0.0146635,0.182578,,0.0720519
95,3894,1,4749,0.0154025,0.190971,,0.0718678
96,3953,1,4847,0.0158898,0.191084,,0.0758952
97,4012,1,4946,0.0164674,0.199609,,0.0754523
98,4071,1,5046,0.0170892,0.204757,,0.0827627
99,4130,1,5147,0.0177393,0.211191,,0.0844313
100,4189,1,5249,0.0184608,0.207741,,0.0857159
101,4248,1,5352,0.0191783,0.223544,,0.0921506
102,4307,1,5456,0.0200149,0.225073,,0.0903847
103,4366,1,5561,0.0206303,0.231928,,0.0905229
104,4425,1,5667,0.0214019,0.232864,,0.0921006
105,4484,1,5774,0.0222084,0.245727,,0.0954088
106,4543,1,5882,0.0229199,0.243195,,0.0974476
107,4602,1,5991,0.0236791,0.248079,,0.0976973
108,4661,1,6101,0.0249039,0.253893,,0.101332
109,4720,1,6212,0.0258001,0.259403,,0.104977
110,4779,1,6324,0.0267034,0.270345,,0.109065
111,4838,1,6437,0.0274573,0.268363,,0.115225
112,4897,1,6551,0.0282445,0.280217,,0.120817
113,4956,1,6666,0.0294969,0.288026,,0.123234
114,5015,1,6782,0.0305458,0.30869,,0.127368
115,5074,1,6899,0.031231,0.298991,,0.12297
116,5133,1,7017,0.0322767,0.312097,,0.125965
117,5192,1,7136,0.0330555,0.313929,,0.129873
118,5251,1,7256,0.0342395,0.319603,,0.131118
119,5310,1,7377,0.0352638,0.319156,,0.133328
120,5369,1,7499,0.0363879,0.329324,,0.138812
121,5428,1,7622,0.0389872,0.339554,,0.136745
122,5487,1,7746,0.0388372,0.345183,,0.142479
123,5546,1,7871,0.0400367,0.349,,0.145362
124,5605,1,7997,0.0414092,0.350976,,0.149228
125,5664,1,8124,0.0432436,0.365713,,0.150029
126,5723,1,8252,0.0439686,0.363854,,0.150447
127,5782,1,8381,0.0454352,0.379809,,0.155599
128,5841,1,8511,0.0465073,0.380206,,0.158133
\end{filecontents*}
\csvreader[range=23-26, no head, late after line=\\, late after last line=\\, separator=comma]{results_qft.csv}{}{%
				\tablenum[table-format=3.0, round-integer-to-decimal=false]{\csvcoli} & 
				\tablenum[table-format=4.0, round-integer-to-decimal=false]{\csvcolii} & 
				\tablenum[table-format=1.0, round-integer-to-decimal=false]{\csvcoliii} & 
				\tablenum[table-format=4.0, round-integer-to-decimal=false]{\csvcoliv} & 
				\tablenum[table-format=1.4, round-mode = places,round-precision = 4, scientific-notation=fixed, fixed-exponent=0]{\csvcolv} & 
				\tablenum[table-format=3.3, round-mode = places,round-precision = 3, group-minimum-digits = 4]{\csvcolvi} &
				\tablenum[table-format=3.4, round-mode = places,round-precision = 4, scientific-notation=fixed, fixed-exponent=0]{\csvcolvii} & 
				\tablenum[table-format=1.3, round-mode = places,round-precision = 3]{\csvcolviii}}\midrule
			\csvreader[range=125-128, no head, late after line=\\, late after last line=\\\bottomrule, separator=comma]{results_qft.csv}{}{%
				\tablenum[table-format=3.0, round-integer-to-decimal=false]{\csvcoli} & 
				\tablenum[table-format=4.0, round-integer-to-decimal=false]{\csvcolii} & 
				\tablenum[table-format=1.0, round-integer-to-decimal=false]{\csvcoliii} & 
				\tablenum[table-format=4.0, round-integer-to-decimal=false]{\csvcoliv} & 
				\tablenum[table-format=1.4, round-mode = places,round-precision = 4, scientific-notation=fixed, fixed-exponent=0]{\csvcolv} & 
				\tablenum[table-format=3.3, round-mode = places,round-precision = 3, group-minimum-digits = 4]{\csvcolvi} &
				--- & 
				\tablenum[table-format=1.3, round-mode = places,round-precision = 3]{\csvcolviii}}\\
			\multicolumn{8}{c}{Quantum Phase Estimation} \\\bottomrule	
		\begin{filecontents*}{results_qpe.csv}
2,4,2,5,2.43E-06,2.22E-05,4.07E-06,1.11E-05
3,8,2,11,5.13E-06,1.85E-05,7.27E-06,1.37E-05
4,12,2,18,4.29E-06,3.38E-05,9.38E-06,2.12E-05
5,17,2,26,5.15E-06,6.56E-05,1.03E-05,3.30E-05
6,26,2,35,7.22E-06,0.000115357,1.63E-05,5.44E-05
7,33,2,45,9.78E-06,0.000154286,1.85E-05,7.38E-05
8,41,2,56,1.22E-05,0.000198596,1.88E-05,9.91E-05
9,53,2,68,1.53E-05,0.000298313,2.61E-05,0.000140306
10,63,2,81,1.99E-05,0.00042472,2.76E-05,0.000177916
11,76,2,95,2.40E-05,0.000584386,3.58E-05,0.000226815
12,89,2,110,3.06E-05,0.000673453,4.21E-05,0.000285003
13,102,2,126,3.64E-05,0.000907018,4.58E-05,0.000352041
14,118,2,143,4.28E-05,0.00116971,5.54E-05,0.00044201
15,134,2,161,5.22E-05,0.00146653,6.41E-05,0.000524737
16,150,2,180,6.27E-05,0.00164831,7.18E-05,0.000646005
17,169,2,200,9.24E-05,0.00228213,7.70E-05,0.000742182
18,187,2,221,8.58E-05,0.00244906,8.03E-05,0.000882099
19,206,2,243,9.58E-05,0.00282265,8.16E-05,0.00100571
20,226,2,266,0.000111697,0.00321972,8.35E-05,0.0011342
21,251,2,290,0.000123557,0.00373392,9.96E-05,0.00133906
22,274,2,315,0.000141947,0.0047336,0.000110917,0.00151831
23,297,2,341,0.000168166,0.00529891,0.000115427,0.00169749
24,321,2,368,0.000182925,0.0056264,0.000116987,0.00193111
25,349,2,396,0.000208205,0.00636957,0.000132137,0.00216104
26,375,2,425,0.000236454,0.00720181,0.000194765,0.00257233
27,402,2,455,0.000441039,0.0107531,0.000195455,0.00272448
28,430,2,486,0.000294723,0.0109276,0.000143817,0.00305078
29,463,2,518,0.000324323,0.0108168,0.000163326,0.00341191
30,494,2,551,0.000357442,0.0114845,0.000181266,0.00379047
31,526,2,585,0.00039235,0.0121711,0.000200965,0.00413836
32,558,2,620,0.00043709,0.0132011,0.000203445,0.0044907
33,591,2,656,0.000451209,0.014238,0.000213695,0.00503092
34,628,2,693,0.000530267,0.0157265,0.000236904,0.00548923
35,664,2,731,0.000528018,0.0170948,0.000258194,0.00627221
36,701,2,770,0.000582806,0.0207818,0.000274493,0.00642414
37,739,2,810,0.000646915,0.0239314,0.000302223,0.00704067
38,777,2,851,0.000690854,0.0302303,0.000318632,0.00777647
39,818,2,893,0.000802211,0.0332336,0.000341652,0.00831508
40,859,2,936,0.00086932,0.0369713,0.000362001,0.00882444
41,900,2,980,0.000938897,0.0393308,0.000366741,0.00935534
42,944,2,1025,0.00102501,0.0567218,0.000393261,0.0100536
43,988,2,1071,0.00112749,0.100881,0.00042296,0.0108386
44,1033,2,1118,0.00123713,0.252084,0.000448879,0.011504
45,1079,2,1166,0.00128424,0.674025,0.000492058,0.0132748
46,1125,2,1215,0.00138772,3.04469,0.000479479,0.0131008
47,1173,2,1265,0.00148183,8.30648,0.000529478,0.0143613
48,1220,2,1316,0.00155629,19.7198,0.000555116,0.0155658
49,1266,2,1368,0.00158692,71.9367,0.000551157,0.0159641
50,1314,2,1421,0.00171768,173.294,0.000578106,0.0168364
\end{filecontents*}
\csvreader[range=42-49, no head, late after line=\\, late after last line=\\\bottomrule, separator=comma]{results_qpe.csv}{}{%
		\tablenum[table-format=3.0, round-integer-to-decimal=false]{\csvcoli} & 
		\tablenum[table-format=4.0, round-integer-to-decimal=false]{\csvcolii} & 
		\tablenum[table-format=1.0, round-integer-to-decimal=false]{\csvcoliii} & 
		\tablenum[table-format=4.0, round-integer-to-decimal=false]{\csvcoliv} & 
		\tablenum[table-format=1.4, round-mode = places,round-precision = 4, scientific-notation=fixed, fixed-exponent=0]{\csvcolv} & 
		\tablenum[table-format=3.3, round-mode = places,round-precision = 3, group-minimum-digits = 4]{\csvcolvi} &
		\tablenum[table-format=3.4, round-mode = places,round-precision = 4, scientific-notation=fixed, fixed-exponent=0]{\csvcolvii} & 
		\tablenum[table-format=1.3, round-mode = places,round-precision = 3]{\csvcolviii}}		
	\end{tabular}}\\\vspace{1mm}
{\scriptsize $n$: Number of qubits \hspace*{0.4cm} \emph{$\vert G \vert$}: Number of gates \\ $t_{\mathit{trans}}$: Runtime of the transformation scheme from \autoref{sec:circuit_transform} \\$t_{\mathit{ver}}$: Runtime of the subsequent equivalence check \\ $t_{\mathit{extract}}$: Runtime of the scheme from \autoref{sec:density_sim} for dynamic circuit \\$t_{\mathit{sim}}$: Runtime of classical simulation for static circuit}\vspace*{-1mm}
\end{table*}

In a first series of evaluations, we employ the scheme proposed in \autoref{sec:circuit_transform} to eliminate the non-unitaries from the dynamic circuit and, afterwards, apply the generic \enquote{proportional} strategy of QCEC for checking the equivalence of the resulting circuit with the corresponding original circuit.
As can be seen from the results, transforming the dynamic circuit using the proposed scheme incurs practically no overhead ($t_\mathit{trans}$ is on the order of \SI{1}{\milli\second} for all tested instances) and allows to successfully verify the full functional equivalence of the Bernstein-Vazirani and QFT algorithms with up to $128$ qubits in a fraction of a second. Even the QPE instances with up to $50$ qubits can be verified in less than \SI{3}{\minute}.

In a second series of evaluations, we use the iterative simulation scheme proposed in \autoref{sec:density_sim} (without employing paralellization) to extract the distribution of measurement outcomes from the dynamic circuits.
In addition, we classically simulate the static counterpart in order to judge the runtime overhead of the proposed scheme.
As expected from the discussion at the beginning of \autoref{sec:density_sim}, only checking the equivalence for a fixed input is an easier task compared to checking the full functional equivalence, in general.
The results for the Bernstein-Vazirani and QPE algorithm show, that extracting the complete measurement probabilities of a dynamic circuit can, in fact, be faster than classically simulating the static counterpart by more than an order of magnitude.
This is in line with the discussions at the end of \autoref{sec:density_sim} and can be attributed to the fact, that the respectively resulting state vectors are sparse, i.e., feature only few non-zero amplitudes.
In contrast, the state vector resulting from the Quantum Fourier Transform is dense, which immediately reflects in the runtime of the extraction scheme, i.e., it roughly doubles with every added qubit.
Thus, the scheme from \autoref{sec:circuit_transform} should be preferred in this case.

\section{Conclusions}\label{sec:conclusions}

In this work, we discussed the upcoming challenges that are currently emerging with the introduction of dynamic quantum circuits. 
We showed that, due to their non-unitary nature, most existing solutions cannot be used for these circuits anymore in an \mbox{out-of-the-box} fashion.
Afterwards, we presented dedicated schemes that eventually allow to treat the involved circuits as if they did not contain non-unitaries at all.

More precisely, the usage of established verification techniques for dynamic quantum circuits is enabled by handling non-unitaries either through
\begin{enumerate}
	\item transforming the dynamic circuit primitives by substituting reset operations with \enquote{fresh} qubits and applying the deferred measurement principle (see \autoref{sec:circuit_transform}), or 
	\item using classical simulation techniques to extract the distribution of measurement outcomes from a dynamic quantum circuit---given a fixed input (see \autoref{sec:density_sim}).
\end{enumerate}
These schemes form a generic solution for handling non-unitaries in verifying the equivalence of quantum circuits that, as demonstrated in our evaluations, is applicable to any existing verification framework.

\section*{Acknowledgments}
%The authors want to thank D. Lummerstorfer for his work on the implementation of the proposed methods.

This project has received funding from the European Research Council (ERC) under the European Union’s Horizon 2020 research and innovation programme (grant agreement No. 101001318).
It has partially been supported by the LIT Secure and Correct Systems Lab funded by the State of Upper Austria as well as by the BMK, BMDW, and the State of Upper Austria in the frame of the COMET program (managed by the FFG).

\clearpage
\balance
\printbibliography
\end{document}